 \documentclass[aps,pre,reprint,groupedaddress,twoside]{revtex4-1}

\usepackage{graphicx}
\usepackage{dcolumn}
\usepackage{bm}
\usepackage{amsmath,amssymb}
\usepackage{color}

\usepackage[absolute]{textpos}
\begin{document}
%
\title{Quantifying uncertainty in state and parameter estimation}

\author{Ulrich Parlitz, Jan Schumann-Bischoff, and Stefan Luther}    
\affiliation{Max Planck Institute for Dynamics and Self-Organization\\ Am Fa\ss berg 17, 37077 G\"ottingen, Germany }
\affiliation{Institute for Nonlinear Dynamics, Georg-August-Universit\"at G\"ottingen, \\  Am Fa\ss berg 17, 37077 G\"ottingen, Germany }

\date{\today}

\begin{abstract}
Observability of state variables and parameters of a  dynamical system from an 
observed time series is analyzed and quantified by means of the Jacobian matrix 
of the delay coordinates map. 
For each state variable and each parameter to be estimated a measure of 
uncertainty is introduced depending on the current state and parameter values, 
which allows us to identify regions in state and parameter space where the 
specific unknown quantity can (not) be estimated from a given time series. 
The method is demonstrated using the Ikeda map and the  Hindmarsh-Rose model.
\end{abstract}

\maketitle
\begin{textblock}{180}(18,262)
\noindent \small{Copyright 2014 by the American Physical Society. The following 
article appeared in U. Parlitz, \textit{et al.}, Phys. Rev. E \textbf{89}, 
050902(R) (2014) and may be found at 
\url{http://dx.doi.org/10.1103/PhysRevE.89.050902}.}
\end{textblock}
%
In physics and other fields of science including quantitative biology, life 
sciences, and climatology, mathematical models play a crucial role for 
understanding and predicting dynamical processes. 
In the following we assume that such a model exists and is known. But even in 
the ideal case of a model obtained from fundamental physical laws this model 
typically contains some parameters whose values have to be determined depending 
on the physical context. Furthermore, not all state variables of the model may 
be easily experimentally accessible. To estimate the unknown parameters and 
state variables you may either devise specific experiments focusing on the 
quantity of interest or you can try to extract the required information from a 
measured time series of the process to be modeled. Technically, several 
estimation methods exist, including observer or synchronization schemes 
\cite{PJK96,NM97,HLN01,GB08,ACJ08,SO09}, particle filters \cite{L10}, a path 
integral formalism \cite{A09,QA10}, or optimization based algorithms 
\cite{CGA08,B10,SBP11}. However, these methods may fail and at this point the 
question arises whether the failure is due to the specific algorithm used or due 
to a lack of information in the available time series. 
In this article we address the second option and present a general approach for 
answering the question  whether a given time series enables the estimation of 
parameters or variables of interest in a given model. The mathematical tool that 
is used to answer this question is delay reconstruction 
\cite{Aeyels,Takens,SYC91,KS97,book_HDIA} and the basic criterion for local 
observability is the rank of the Jacobian matrix of the delay coordinates map. 
This approach was motivated by work of Letellier, Aguirre, and Maquet 
\cite{LAM05,LA09,FBL12} who studied the question which state variables can be 
estimated or observed from a given time series using derivative coordinates. 
Observability of (continuous) dynamical system  is also a major issue in control 
theory \cite{HK77,Sontag, N82} and nonlinear time series analysis \cite{VTK04}. 
Here we consider discrete time and delay coordinates, and we introduce a 
quantitative measure of uncertainty which in general varies on the attractor and 
thus indicates where in state space estimation is more efficient and less error 
prone.  Furthermore, we focus not only on state variables but also on 
observability of model parameters.

Let's assume, first, that our model of interest is  a $M$-dimensional discrete 
dynamical system
\begin{equation}  \label{discrsyst}
   {\mathbf x}(n+1) = {\mathbf g} [ {\mathbf x}(n), \mathbf{p}] 
\end{equation}
given by an iterated function $\mathbf{g}$ depending on the state vector 
$\mathbf{x}(n) = (x_1(n),\ldots,x_M(n)) \in \mathbb{R}^M$ at time $n$ and $K$ 
parameters $\mathbf{p} = (p_1, \ldots, p_K) \in \mathbb{R}^K$. This system 
generates the times series $\{ s(n) \}$ with $s(n) = h( {\bf x}(n))$ (for $n = 
1,\ldots, N$), where $h$ denotes a measurement or observation function. The time 
series $\{ s(n) \}$ can be used to construct a $D$ dimensional {\em delay 
reconstruction} \cite{Aeyels,Takens,SYC91,KS97,book_HDIA},
\begin{eqnarray} \label{delemb}
  {\mathbf y} & = & \left(s(n), s(n+1), ....,  s(n+D-1) \right)    \\ \nonumber
           & = &  G( {\mathbf x}, \mathbf{p}) \in \mathbb{R}^D
\end{eqnarray}
providing the {\em delay coordinates map}  $G: \mathbb{R}^{M +K} \to \mathbb{R}^D$. 

To uniquely recover  the full state $\mathbf{x}$ and the parameters $\mathbf{p}$ 
from the observations represented by the reconstructed state $\mathbf{y}$ the 
map $G$ has to be locally invertible. More precisely, let $M+K \le D$ and let $ 
(\mathbf{x}, \mathbf{p}) \in {\cal{U}}$ where ${\cal{U}} \subset 
\mathbb{R}^{M+K}$ is a smooth manifold. Then $G$ is locally invertible on the 
image $G({\cal{U}})  \subset  \mathbb{R}^D$ if the  $D \times (M+K)$ Jacobian 
matrix $DG(\mathbf{x}, \mathbf{p})$  has full rank $M+K$ (i.e., $G$ is an 
immersion \cite{SYC91}).

The map from delay reconstruction space $\mathbb{R}^D$ to the state and 
parameter space $\mathbb{R}^{M+K}$  is locally given by the (pseudo) inverse of 
the Jacobian matrix $DG$ of the delay coordinates map $G$, which can be computed 
using a singular value decomposition 
\begin{equation} \label{SVD1}
   DG = U S V^\text{tr}
\end{equation}
where $S = \rm{diag}( \sigma_1, \ldots , \sigma_{M+K}) $ is a $(M+K) \times 
(M+K)$ diagonal matrix containing the singular values $\sigma_1 \ge \sigma_2 \ge 
\ldots \ge \sigma_{M+K} \ge 0$  and $U = (\mathbf{u}^{(1)}, \ldots , 
\mathbf{u}^{(M+K)}) $ and $V = ( \mathbf{v}^{(1)}, \ldots , \mathbf{v}^{(M+K)} ) 
$ are orthogonal matrices, represented by the column vectors $\mathbf{u}^{(i)} 
\in \mathbb{R}^D$ and $\mathbf{v}^{(i)} \in \mathbb{R}^{M+K}$, respectively. 
$V^\text{tr}$ is the transposed of $V$ coinciding with the inverse $V^{-1} = 
V^\text{tr}$. Analogously, $U^\text{tr} = U^{-1} $ and the (pseudo) inverse 
Jacobian matrix reads $DG^{-1} =  V S^{-1} U^\text{tr}$ where 
$S^{-1} = \rm{diag}(1/\sigma_1, \ldots, 1/ \sigma_{M+K}) $. Multiplying by $U$ 
from the right we obtain $ DG^{-1} U =  V S^{-1}  $ or
\begin{equation} \label{SVD3}
   DG^{-1}  \mathbf{u}^{(j)} = \frac{1}{\sigma_j} \mathbf{v}^{(j)}  \  \  \  \  \  (j = 1, \ldots, M+K) .
\end{equation}

\begin{figure} 
\centering
\includegraphics[width=8.5cm]{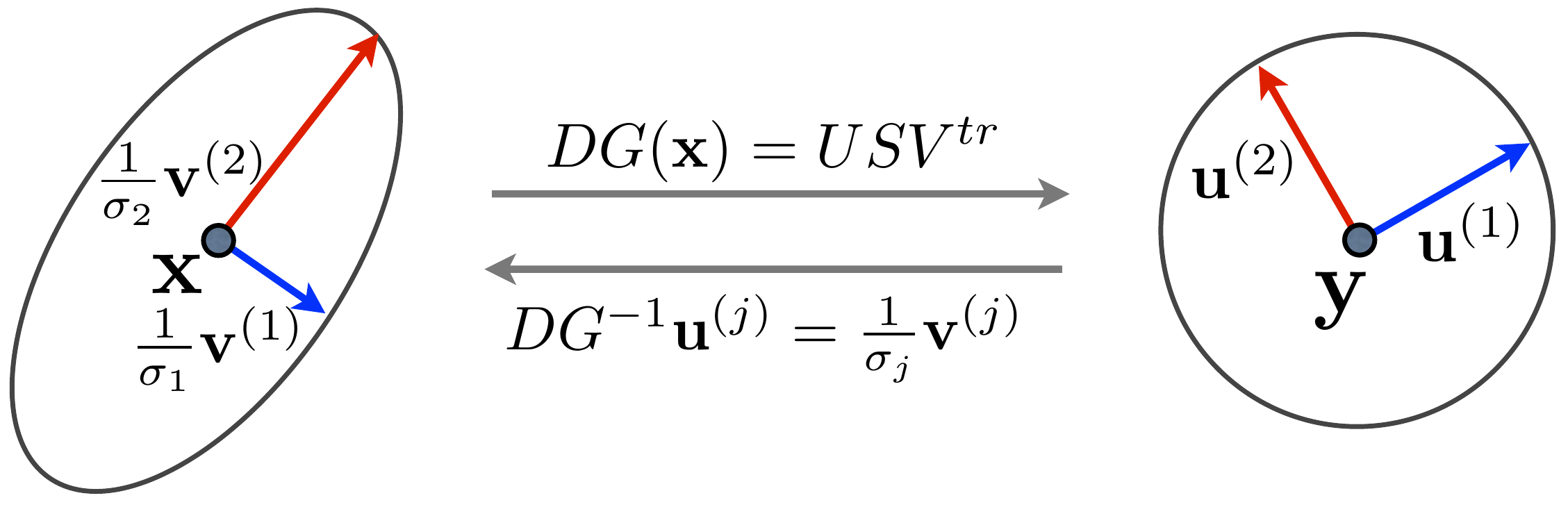}
\caption{(Color online) The (pseudo) inverse Jacobian matrix 
        $DG^{-1}(\mathbf{y})$ maps perturbations of $\mathbf{y}$ 
        in delay reconstruction  space to deviations from the state 
        $\mathbf{x}$ whose magnitudes depend on the direction 
        of the perturbation as described by Eq.~(\ref{SVD3}). }
\label{figB}
\end{figure}

In Fig.~\ref{figB} the transformation of singular vectors Eq. \eqref{SVD3} is 
illustrated for the case $M=2$ and $K=0$ (no unknown parameters).  The diagram 
shows how small perturbations of $\mathbf{y}$ in delay reconstruction space 
result in deviations from $\mathbf{x}$ in the original state space. 
Most relevant for the local observability of the (original) state $\mathbf{x}$ 
is the length of the longest principal axis of the ellipsoid given by the 
inverse of the smallest singular value $\sigma_2$ (see Fig.~\ref{figB}). 
Small singular values correspond to directions in state space, where it is 
difficult (or even impossible) to locate the true state $\mathbf{x}$ given a 
finite precision of the reconstructed state $\mathbf{y}$. 
The ratio $\sigma_\text{min} / \sigma_\text{max}$  of the smallest and the 
largest singular value is a measure of observability at the reference state 
$\mathbf{x}$. By averaging on the attractor we define (analogously to a similar 
definition for derivative coordinates \cite{LAM05,LA09}) the 
\textit{observability index}
\begin{equation}  \label{obsindx}
   \bar \gamma = \frac{1}{N}  \sum_{n=1}^N 
\frac{\sigma_\text{min}^2(\mathbf{x})} { \sigma_\text{max}^2 (\mathbf{x} )}.
\end{equation}
%

%
%

If the perturbations of $\mathbf{y}$ are due to normally distributed measurement 
noise than they can be described by a symmetric Gaussian distribution centered 
at $\mathbf{y}$
\begin{equation}
    Q(\mathbf{\tilde y}) = \frac{ \exp \left[ -\frac{1}{2} ({\mathbf{\tilde y}} 
- {\mathbf{y}})^\text{tr} \Sigma_y^{-1} ({\mathbf{\tilde y}} - {\mathbf{y}}) 
\right]
   }{ \sqrt{  (2\pi)^D \det (\Sigma_y)  }   }
\end{equation}
where $\mathbf{\tilde y}$ is the perturbed state, $\Sigma_y = 
{\rm{diag}}(\rho^2, \ldots, \rho^2) = \rho^2 I_D $ denotes the $D \times D$ 
covariance matrix ($I_D$ stands for the $D$-dimensional unit matrix), and the 
standard deviation $\rho$ quantifies the noise amplitude. For (infinitesimally) 
small perturbations $\mathbf{\Delta y}= \mathbf{\tilde y} - \mathbf{y} $ this 
distribution is mapped by the pseudo inverse of the linearized delay coordinates 
map to the (non-symmetrical) distribution 
\begin{equation}
   P(\mathbf{\tilde x}) = \frac{\exp \left[ -\frac{1}{2} (\mathbf{\tilde x} - 
\mathbf{x})^\text{tr} \Sigma_x^{-1} (\mathbf{\tilde x} - \mathbf{x}) \right]}{ 
\sqrt{  (2\pi)^{M+K} \det (\Sigma_x)  } } 
    \end{equation}
centered at $\mathbf{x}$ with the inverse covariance matrix
\begin{equation} \label{sigma_x_inv}
   \begin{array} {rcl}
        \Sigma_x^{-1}  & = &  DG^\text{tr} \Sigma_y^{-1} DG  \\
                           & = & \frac{1}{\rho^2}  DG^\text{tr} DG  =   
        \frac{1}{\rho^2} V S^2 V^\text{tr} . 
   \end{array}
\end{equation} 

The marginal distribution $P_j$ of the $j$th state variable centered a $x_j$ 
is given by
\begin{equation}
  P_j( \tilde  x_j) = \frac{1}{\rho_j  \sqrt{2 \pi} }  \exp \left[ - 
  \frac{(\tilde x_j - x_j)^2} {2 \rho_j^2}  \right] \, ,
\end{equation}
where the standard deviation $\rho_j$ is given by the square root of the 
diagonal elements of the covariance matrix %
   $ \rho_j = \sqrt{ \Sigma_{x,jj} }  $
%
that can be obtained by inverting $\Sigma_x^{-1}$ [given in 
Eq.~(\ref{sigma_x_inv})]. Since the noise level $\rho$ of the observations 
appears in Eq.~(\ref{sigma_x_inv}) as a factor only we can, without loss of 
generality, choose $\rho = 1$ and use
\begin{equation} \label{uncert}
    \nu_j = \sqrt{ \left[ DG^\text{tr}  DG  \right]^{-1}_{jj} }   =  
    \sqrt{\left[ V S^{-2} V^\text{tr} \right]_{jj} }   
\end{equation}
as a  measure of \textit{uncertainty} when estimating $x_j$, which can be 
interpreted as a noise amplification factor. The same reasoning holds for the 
unknown parameters $\mathbf{p}$.

\begin{figure} 
\centering
\includegraphics{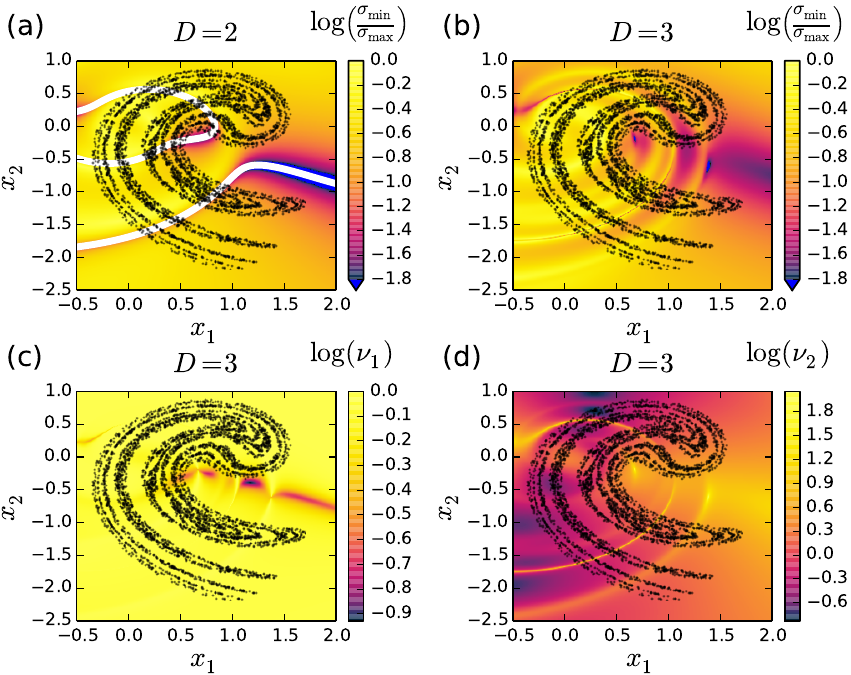}
\caption{(Color online) Observability of the state variables $x_1$ and $x_2$ 
        of the Ikeda map
        Eq.~\eqref{ikeda_map} from a  $x_1$ time series (with known parameters, 
        $K=0$, $M=2$). (a), (b)  Color-coded ratio of singular values 
        $\sigma_\text{min} /  \sigma_\text{max}$ vs. $x_1$ and $x_2$ 
        for reconstruction dimension  $D=2$ (a) and $D=3$ (b). The white curves 
        in (a) indicate the location of zeros of $\det(DG)$. (c), (d)
        Color-coded uncertainties $\nu_1$ (c) and $\nu_2$ (d) of $x_1$ and $x_2$ 
        estimates, respectively. Note the logarithmic color axes. Black dots 
        represent the Ikeda attractor.}             
\label{fig2}
\end{figure}

To illustrate this quantification of observability we first consider the Ikeda 
map \cite{I79} 
$z(n+1) = p_1 + p_2 z(n) \exp[ i p_3 - i p_4 / (1 + \vert z(n) \vert^2)   ] $
with $z(n) = x_1(n) + i x_2(n)  \in \mathbb{C}$ that can also be written as
\begin{equation} \label{ikeda_map}
    \begin{array} {rcl}
   x_1(n+1) & = & p_1 + p_2  [x_1(n) \cos \theta_n - x_2(n) \sin \theta_n] \\ 
   x_2(n+1) & = &          p_2 [x_1(n) \sin \theta_n + x_2(n) \cos \theta_n] 
   \end{array}
\end{equation}
where $\theta_n = p_3 - p_4/[1 + x_1^2(n) + x_2^2(n) ]  $.
For the standard parameters $p_1=1$, $p_2=0.9$, $p_3=0.4$, and $p_4=6$ this map 
generates the chaotic attractor shown in Fig.~\ref{fig2}.

First, we consider a  case where all parameters are known and only the variables 
$x_1$ and $x_2$ have to be estimated from the observable $s(n) = x_1(n)$ (i.e., 
$M=2$ and $K=0$). Figures~\ref{fig2}(a) and \ref{fig2}(b) show (color-coded) the 
ratio of the smallest singular value $\sigma_\text{min}=\sigma_M$ and the 
largest singular value $\sigma_\text{max}=\sigma_1$ of the Jacobian matrix 
$DG(\mathbf{x})$ of the delay coordinates map vs. $x_1$ and $x_2$. 
Reconstruction dimensions are $D=2$ in Fig.~\ref{fig2}(a) and $D=3$ in 
Fig.~\ref{fig2}(b), respectively. For $D=2$, the white curves indicate the zeros 
of the determinant of $DG(\mathbf{x},\mathbf{p})$ that are computed as contour 
lines.
As can be seen parts of the Ikeda attractor  cross these singularity manifolds 
or are close to regions in state space where the ratio $\sigma_\text{min} / 
\sigma_\text{max}$ is very close to zero, indicating an almost singular Jacobian 
matrix $DG$. There, state estimation is not possible, a fact that reconfirms 
previous results indicating that reconstruction dimensions $D>2$ are required 
for the Ikeda map \cite{BBA90}.
For $D=3$ the singularities disappear and only some regions with relatively low 
ratios $\sigma_\text{min} / \sigma_\text{max}$ remain.

Figures~\ref{fig2}(c) and \ref{fig2}(d) show $\nu_1$ and $\nu_2$ versus $x_1$ 
and $x_2$, respectively. For both variables their uncertainties $\nu_k$ vary 
and there are regions of low $\nu_1$ but relatively large $\nu_2$.
 
\begin{figure} 
\centering
\includegraphics[]{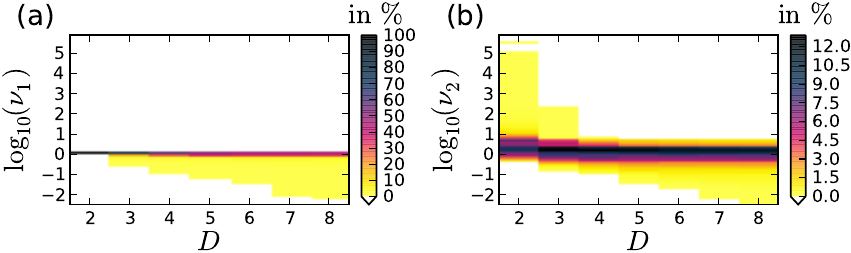}
\caption{(Color online) Histograms (color-coded) of uncertainties $\nu_1$ (a) 
        and $\nu_2$ (b) computed from a $x_1$ time series of length $N=1000000$ 
        generated on the attractor of the Ikeda map Eq.~\eqref{ikeda_map} with 
        reconstruction dimensions ranging from $D=2$ to $D=7$. The state 
        variables $x_1$ and $x_2$ are estimated ($M=2$) while all parameters are 
        assumed to be known ($K=0$).}
\label{fig3}
\end{figure}

Figures~\ref{fig3}(a) and \ref{fig3}(b) show histograms of $\nu_1$ and $\nu_2$ 
for different reconstruction dimensions $D$ which were obtained from an orbit of 
length $N=1000000 $ on the Ikeda attractor. Due to the choice $s(n) = x_1(n)$ 
the uncertainty $\nu_1$ of $x_1$ is for all dimensions equal or less than one.  
For $D=2$ the uncertainty $\nu_2$ of $x_2$ reaches very high values $ > 10^6$ 
when the orbit passes those regions in state space where the Jacobian matrix 
$DG$ is (almost) singular [see Fig.~\ref{fig2}(a)]. For reconstruction 
dimensions $D=3$ the $\nu_2$-histogram is bounded by $\nu_2 < 10^3$ indicating a 
significant improvement and for  $D = 4$ the bound reduces to $\nu_2 < 10$, a 
value that doesn't change anymore if the reconstruction dimension is increased 
furthermore. This feature is in very good  agreement with previous results 
obtained when estimating Lyapunov exponents from Ikeda time series \cite{BBA90}.

To obtain the histograms shown in Fig.~\ref{fig3} and in the following figures 
the model equations are used to generate a trajectory which provides a 
representative sample and subset of the attractor (similar to numerical 
computations of Lyapunov exponents). 

\begin{figure} 
\centering
\includegraphics[]{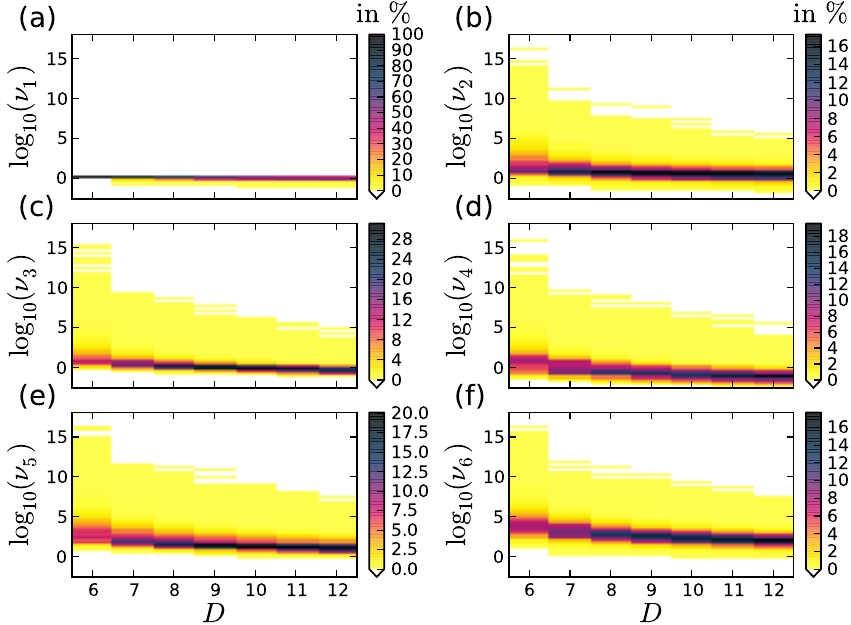}
\caption{(Color online) Histograms (color-coded) of uncertainties of state and 
        parameter estimates of the Ikeda map  Eq.~\eqref{ikeda_map} for 
        reconstruction dimensions ranging from $D=6$ to $D=12$. Distributions 
        are computed from a $x_1$ time series of length $N=1000000$ generated on 
        the Ikeda attractor. All variables ($M=2$) and all parameters ($K=4$) 
        are assumed to be unknown. }
\label{fig4}
\end{figure}

For the results shown in Figs.~\ref{fig3}(a) and \ref{fig3}(b) only the state 
variables are estimated and all parameters are assumed to be known  ($M=2$, 
$K=0$). Figure \ref{fig4} shows also the uncertainties $\nu_3$, $\nu_4$, 
$\nu_5$, and $\nu_6$ of the parameters $p_1$, $p_2$, $p_3$, and $p_4$ for an 
estimation task where all variables ($M=2$) and all parameters ($K=4$) are 
unknown. For increasing reconstruction  dimension $D$ the distributions of all 
uncertainties converge with monotonically decreasing upper bounds (largest 
$\nu$-values quantifying large uncertainty of estimates at specific locations on 
the attractor).

Delay reconstruction can also be applied to observables $s(t) = 
h[\mathbf{x}(t)]$ from continuous dynamical systems,
\begin{equation}  \label{contsyst}
   \mathbf{ \dot x} = \mathbf{f}( \mathbf{x}, \mathbf{p} ) \, ,
\end{equation}
using a suitable {\em delay time} $\tau$:
\begin{equation} \nonumber 
      {\mathbf y}  =  \left(s(t), s(t+\tau), ....,  s(t+(D-1)\tau)    \right)   
                =  G( {\mathbf x}, \mathbf{p}) \in \mathbb{R}^{D} .
 \end{equation}
The Jacobian matrix $DG({\mathbf x}, \mathbf{p})$ of the delay coordinates map 
$G$  can be computed by solving linearized equations providing the Jacobian 
matrices $D_x\phi^t  (\mathbf{x},\mathbf{p})$  and $D_p\phi^t  
(\mathbf{x},\mathbf{p})$ of  the flow $\phi^t$ generated by the system 
Eq.~\eqref{contsyst} \cite{Kawakami}. To demonstrate the application of the 
proposed uncertainty analysis to continuous time system we use the 
Hindmarsh-Rose (HR) neuron model \cite{HR84}
\begin{eqnarray}
	\dot x_1 	&=&	-x_1^3+p_1 x_1^2+x_2-x_3 \nonumber \\
	\dot x_2 	&=&	1-p_2  x_1^2-x_2    \label{HRmod} \\
	\dot x_3 	&=&	p_3  \left( x_1 + p_4  (p_5-x_3) \right) .  \nonumber
\end{eqnarray}
For parameter values $p_1=3$, $p_2 = 5$, $p_3=0.004$,  $p_4 =3.19$, $p_5=0.25$ 
the HR model exhibits chaotic bursting of $x_1$ and $x_2$ and slow variations of 
$x_3$ \cite{SBP11}.

%
\begin{figure} 
\centering
\includegraphics[]{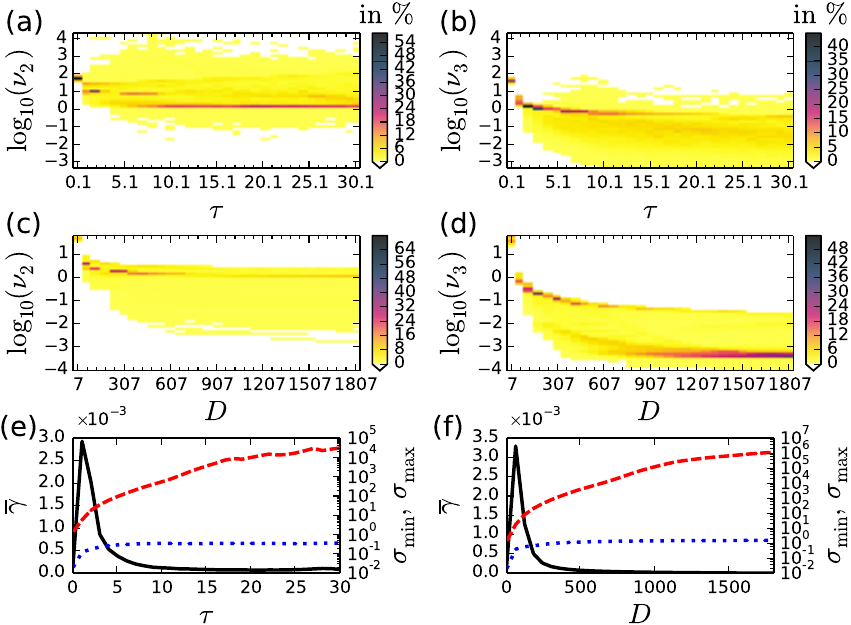}
\caption{(Color online) Probability distributions (color-coded) of         
        uncertainties 
        $\nu_2$ and $\nu_3$ when estimating the state variables $x_1$, $x_2$, 
        and $x_3$  of the HR-model Eq.~\eqref{HRmod} from a $x_1$ time series.
        In (a) and (b) the delay reconstruction dimension is fixed at $D=7$  and 
        the delay $\tau$ is varied. (c), (d) show distributions for $\tau=0.1$ 
        and different reconstruction dimensions $D$. Corresponding columns 
        (histograms) of all four diagrams show results for the same window in 
        time $(D-1) \tau$ used upon delay reconstruction. (e), (f) 
        Observability index $\bar \gamma$ \eqref{obsindx} (solid curve), 
        $\sigma_\text{min}$ (dotted curve), and $\sigma_\text{max}$ (dashed 
        curve)  vs. $\tau$ and vs. $D$.
}
\label{fig5}
\end{figure}

Figures~\ref{fig5}(a) and \ref{fig5}(b) show the dependence of probability 
distributions (color-coded) of uncertainties $\nu_2$, and $\nu_3$, respectively, 
 on the delay  time  $\tau$ chosen for performing the delay reconstruction. The 
reconstruction dimension equals $D=7$. With this example, all parameters are 
assumed to be known ($K=0$) and the first state variable is chosen as measured  
time series $s(t_n) = x_1(t_n)$ with $t_n = n \tau$. Therefore, the estimation 
of $x_1$ is not much affected by the choice of the delay time and $\nu_1 \le 1 $ 
 (with $\nu_1 \approx 1$ most of the time, not shown here). As can be seen the 
centers of both distributions decrease monotonically with $\tau$ indicating an 
improvement of the estimation accuracy for larger delay times. 
Figures~\ref{fig5}(c) and \ref{fig5}(d) show histograms (color-coded) of 
uncertainties $\nu_2$, and $\nu_3$ versus reconstruction dimension $D$ for $\tau 
= 0.1$. Larger $D$ provides lower uncertainties $\nu_j$ and compared to 
Figs.~\ref{fig5}(a) and  \ref{fig5}(b) very large $\nu_j$ do not occur anymore. 
Note that corresponding columns of Figs.~\ref{fig5}(a) and \ref{fig5}(b) and 
Figs.~\ref{fig5}(c) and \ref{fig5}(d), respectively, are computed using delay 
coordinates covering  the same window in time ranging from $\tau(D-1) = 0.1 
\cdot  6 = 0.6$ to $\tau(D-1)= 30.1 \cdot6 = 1806 \cdot 0.1 = 180.6$. The more 
densely sampling ($\tau = 0.1$) underlying Figs.~\ref{fig5}(c) and \ref{fig5}(d) 
provides more information about the underlying dynamics and results in lower 
uncertainty values. 
Figures~\ref{fig5}(e) and \ref{fig5}(f) show the observability index $\bar 
\gamma$  Eq.~\eqref{obsindx} and mean values of the smallest and the largest 
singular values $\sigma_\text{min}$ and $\sigma_\text{max}$ versus $\tau$ and  
$D$, respectively. While $\bar \gamma$ exhibits a clear peak, 
$\sigma_\text{min}$ converges to an asymptotic value, and $\sigma_\text{max}$ 
increases monotonically, i.e., the lengths of the ellipsoid axes in 
Fig.~\ref{figB}  decrease ($1/\sigma_\text{max}$) or converge 
($1/\sigma_\text{min}$).

\begin{figure} 
\centering
\includegraphics[width=8.4cm]{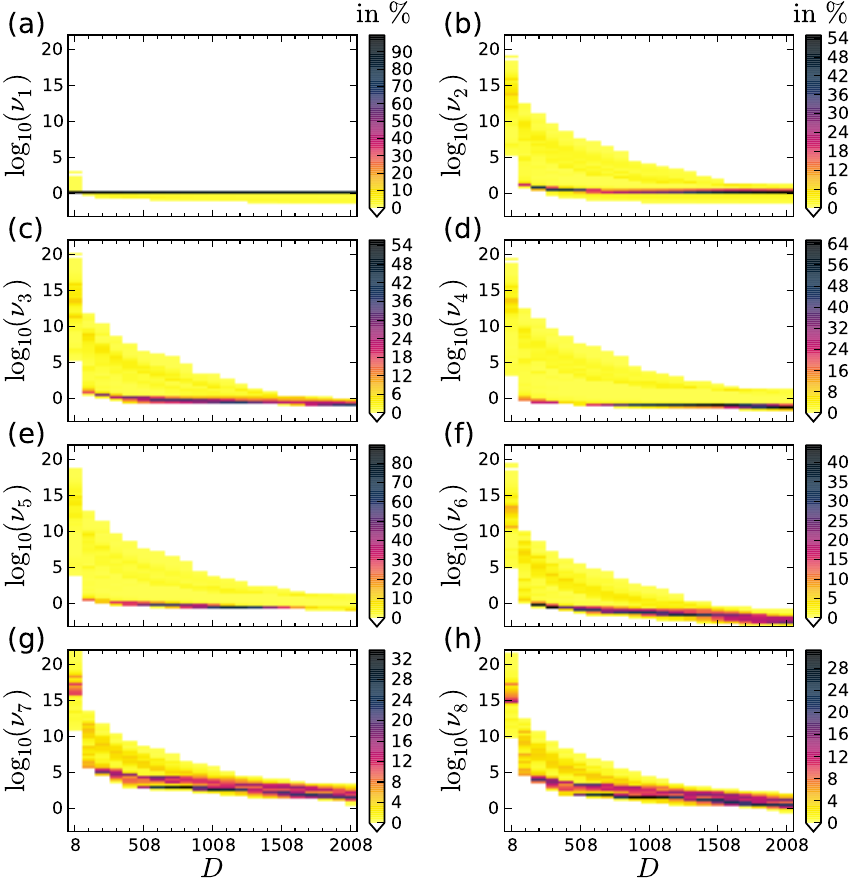}
\caption{(Color online) Distributions of uncertainties $\nu_j$ vs. 
        reconstruction dimension $D$ obtained for the HR model \eqref{HRmod} 
        where all three state variables and all five parameters are estimated 
        from a $x_1$ time series. The delay time $\tau = 0.1$ is fixed.}
\label{fig6}
\end{figure}

If in addition to the three state variables $x_1$, $x_2$, and $x_3$ also the 
five parameters $p_1, \ldots, p_5$ of the HR-model Eq.~\eqref{HRmod} are to be 
estimated from the $x_1$ time series then we have to cope with an estimation 
task with $M+K = 3+5=8$ uncertainties whose distributions for $\tau = 0.1$ are 
shown in Fig.~\ref{fig6} for delay reconstruction dimensions ranging from $D=8$ 
to $D=2008$. For increasing $D$ the uncertainties $\nu_1, \ldots, \nu_6$ 
corresponding to $x_1, x_2, x_3, p_1, p_2, p_3$ decrease to values close to or 
below one. The uncertainties $\nu_7$ and $\nu_8$ of parameters $p_4$  and $p_5$, 
respectively, remain rather large ($> 1000$) even for high dimensional 
reconstructions. This feature indicates that it is very difficult to estimate 
both parameters together. In fact, if $p_4$ (or $p_5$) is known and only $p_5$ 
(or $p_4$) has to be estimated (together with $x_1, x_2, x_3, p_1, p_2, p_3$) 
then the uncertainty values of $p_5$ (or $p_4$) are much smaller and lie in the 
range of the uncertainties of the other parameters. Applying a state and 
parameter  estimation algorithm \cite{SBP11,SBLP13} we also encountered problems 
(in terms of large deviations from the true values) when trying to estimate both 
parameters $p_4$ and $p_5$ together. These two parameters are to some degree 
redundant in the sense that different combinations yield (almost) the same $x_1$ 
time series and thus  cannot be clearly distinguished using a $x_1$ time series, 
only. 


The presented approach for quantifying uncertainties of model based state and 
parameter estimation from time series provides a general criterion whether and 
how reliably specific model variables and parameters can be estimated from  time 
series. This method is independent from any particular estimation method and it 
can be extended in several ways, including unknown parameters in the measurement 
function  and multivariate time series. High uncertainty implies that the 
corresponding quantity of the model has  small impact on the output and may thus 
be a candidate for reducing the formal model complexity by pruning. Furthermore, 
the information provided by the values of uncertainty can be exploited to 
improve state and parameter estimation methods.

\newpage
\begin{acknowledgements}
The research leading to these results has received funding from the European 
Community's Seventh Framework Program FP7/2007-2013 under grant agreement no 
HEALTH-F2-2009-241526, EUTrigTreat. 
We acknowledge financial support by the German Federal Ministry of Education and 
Research (BMBF) Grant No. 031A147, the Deutsche Forschungsgemeinschaft (SFB 
1002: Modulatory Units in Heart Failure), and by the German Center for 
Cardiovascular Research (DZHK e.V.).

\end{acknowledgements}

\end{document}